\documentclass[11pt]{article}

\usepackage[a4paper,lmargin=2.5cm,rmargin=2cm,tmargin=1cm,bmargin=1cm,includehead,includefoot]{geometry}

\usepackage{graphicx,amsmath,url}
\usepackage{microtype,lmodern}
\usepackage{mathtools,amssymb,gensymb}
\usepackage{tgtermes}
\usepackage[T1]{fontenc}
\usepackage[utf8]{inputenc}

\usepackage{acronym}
\usepackage{amsmath}
\usepackage{array}
\usepackage{bbold}
\usepackage{booktabs}
\usepackage{dsfont}
\usepackage{fancyhdr}
\usepackage{relsize}
\usepackage[font={small,up,singlespacing}]{subcaption}
\usepackage[backend=bibtex,style=chem-angew,sorting=none,autocite=superscript]{biblatex}
\usepackage{pdfpages}
\usepackage{bbm}
\usepackage{color}
\usepackage{siunitx}
\usepackage{braket}
\usepackage{mwe}
\usepackage{biblatex}
\usepackage{multirow}
\usepackage{mathtools}
\usepackage{xcolor}
\usepackage{float}
\usepackage{colortbl}
\usepackage{siunitx}\DeclareSIUnit\molar{\textsc{M}}

\usepackage[colorlinks,breaklinks]{hyperref} 
\usepackage{cleveref}

\newcommand{\nz}{$\mathrm{N}_\zeta$ }

\newcommand{\og}{$\mathrm{O}_\gamma$ }
\newcommand{\nzf}{LYS $\mathrm{N}_\zeta$ }
\newcommand{\cdf}{GLU $\mathrm{C}_\delta$ }

\hypersetup{hypertexnames=true, linkcolor=blue, anchorcolor=black, citecolor=blue, urlcolor=blue}
\graphicspath{{./}{imglocal/}{img/}}
\title{\textbf{Implicit Solvent Approach Based on Generalised Born and Transferable Graph Neural Networks for Molecular Dynamics Simulations}}

\author{Paul Katzberger,$^a$ Sereina Riniker$^a$*}
\date{$^a$~Department of Chemistry and Applied Biosciences, ETH Zurich, Vladimir-Prelog-Weg 2, 8093 Zurich, Switzerland. *Email: sriniker@ethz.ch}

\begin{document}

\maketitle

\section*{Abstract}

Molecular dynamics (MD) simulations enable the study of the motion of small and large (bio)molecules and the estimation of their conformational ensembles. The description of the environment (solvent) has thereby a large impact. Implicit solvent representations are efficient but in many cases not accurate enough (especially for polar solvents such as water). More accurate but also computationally more expensive is the explicit treatment of the solvent molecules. Recently, machine learning (ML) has been proposed to bridge the gap and simulate in an implicit manner explicit solvation effects. However, the current approaches rely on prior knowledge of the entire conformational space, limiting their application in practice. Here, we introduce a graph neural network (GNN) based implicit solvent that is capable of describing explicit solvent effects for peptides with different composition than contained in the training set. 

\section{Introduction}
Molecular dynamics (MD) simulations employ Newton's equation of motion to study the dynamics of (bio)molecular systems \cite{Hansson2002Molecularsimulations}. In recent years, MD has not only become a pivotal tool for the investigation of biomolecular processes such as membrane permeation of drug molecules or protein folding \cite{Gruebele2002ProteinSurface} but also accelerated and supported drug discovery \cite{Mortier2015TheComplexes} The conformational ensembles of molecules are strongly influenced by the surrounding medium. While intramolecular hydrogen bonds (H-bonds) are favoured in vacuum and apolar solvents (e.g., chloroform), they are generally disfavoured in polar solvents (e.g., water) \cite{Cumberworth2016FreeModels}. Such effects of the (local) environment can be incorporated by explicitly simulating solvent molecules. This approach provides good accuracy since it includes both short-range and long-range interactions. Both contributions are needed in order to describe an accurate conformational ensemble \cite{Zhang2017ComparisonSolvents}. However, explicit-solvent simulations come at the cost of substantially increasing the number of degrees of freedom (DOF) in the system, which results in substantially higher computational costs as well as slower effective sampling \cite{Roux1999ImplicitModels}. In addition, the potential energy of a single solute conformation is no longer an instantaneous property in explicit-solvent simulations as an infinite number of solvent configurations (i.e., arrangement of the solvent molecules around the solute) exist for a single solute conformation. Thus, the prediction of the potential energy of a single conformer requires integrating out the contributions of individual solvent configurations \cite{Zhang2017ComparisonSolvents}. 

To simultaneously retain the instantaneousity of the potential energy and to reduce the number of DOF in the system, implicit-solvent methods have been developed \cite{Cramer1999ImplicitDynamics}. These approaches aim at modelling the solute-solvent interactions in an implicit manner by predicting the mean solvation energy and forces for a given solute conformation \cite{Roux1999ImplicitModels}. The most common implicit-solvent approach replaces the solvent by a continuum electrostatic. Examples are Poisson-Boltzmann (PB) \cite{Baker2005ImprovingView}, generalised Born (GB) \cite{StillW.Clark1990SemianalyticalDynamics}, or fast analytical continuum treatments of solvation (FACTS) \cite{Haberthur2007FACTS:Solvation}. Note that GB and FACTS models are approximations of the PB model. However, current implicit-solvent models do not accurately reproduce the short-range effects of solvent molecules, and thus often do not reproduce the secondary structures of peptides correctly \cite{J.M.Lang2022GeneralizedPeptides}. Only very recently, the use of machine learning (ML) approaches have started to be explored for implicit-solvent models. Chen \textit{et al.} \cite{Chen2021MachineDynamics} used graph neural networks (GNN) to reproduce the potential energies and forces of two small peptides. This method is, however, not yet practical because the full conformational ensemble needs to be generated first via explicit-solvent simulations, before the GNN model can be trained to reproduce it. The GB models remain therefore the most commonly used implicit solvent approach to date.

The GB equation was first introduced by Still \textit{et al.} \cite{StillW.Clark1990SemianalyticalDynamics} and defines the solvation free energy $\Delta G$ in terms of the atomic partial charges $q_i$ and $q_j$, the distance $r_{ij}$, and the effective Born radii $R_i$ and $R_j$ of two atoms $i$ and $j$. 
\begin{equation}
    \Delta G = -\frac{1}{2} \left ( \frac{1/\epsilon_{in}}{\epsilon_{out}} \right ) \sum_{i,j} \frac{q_iq_j}{\sqrt{r_{ij} + R_iR_j exp \left ( \frac{-r_{ij}^2}{4R_iR_j} \right )}}
\end{equation}
The Born radii are further calculated using the Coulomb integral $I_i$ and the intrinsic radius $\rho_i$,
\begin{equation}
    R_i = (\rho_i^{-1} - I_i)^{-1} .
\end{equation}
The Coulomb integral can be derived from the Coulomb field approximation (CFA) and the intrinsic radius, 
\begin{equation}
    I_i = \frac{1}{4\pi} \int_{\Omega,r>\rho_i} \frac{1}{r^4} d^3 r .
\end{equation}
Typically, the integral is solved analytically by using a pairwise de-screening approximation \cite{Hawkins1995PairwiseMedium}. While the functional form of different GB models is the same, the manner in which this integral is calculated distinguishes them: GB-HCT \cite{Hawkins1995PairwiseMedium}, GB-OBC \cite{Onufriev2004ExploringModel}, or GB-Neck \cite{Mongan2007GeneralizedCorrection}. In addition, also ML has been proposed to directly approximate reference born radii calculated by PB calculations \cite{Horvath2020bigGeometry,MohamedMahmoud2020GeneralizedNetworks}.

The calculation of the effective Born radii in standard GB models can be thought of as a one-pass message passing algorithm, where information is sent from each node to all other nodes within a cutoff. GNNs are therefore a natural choice to develop GB based models further. When multiple passes are used, GNN can aggregate information, intrinsically encode the geometric environment, and thereby go beyond the pairwise de-screening approximation. In general, it is mainly the local environment, dominated by short-range interactions, which is expected to benefit from this description, as the GB models can describe long-range interactions well. Therefore, the robustness and quality of a GNN based implicit solvent could be enhanced by using a $\Delta$-learning \cite{Ramakrishnan2015BigApproach} approach rather than predicting the solvation forces directly with the GNN. This means that rather than replacing the GB model entirely, a ML correction could be added to a base model (i.e., GB-Neck). In related fields such as ML for QM/MM simulations of condensed-phase systems, a similar approach has been demonstrated to lead to stable simulations \cite{Boselt2021MachineSystemsb}. In addition, the $\Delta$-learning scheme could allow smaller cutoffs for the GNN as the long-range interactions are already well described by the GB model, reducing the computational cost.

Here, we use this idea to develop a ML based implicit solvent approach, which can be used to simulate molecules without the need of a complete conformational ensemble from explicit-solvent simulations for training. The training set for our ML approach consists of subsets of conformers extracted from explicit-solvent simulations (here with the TIP5P water model \cite{Khalak2018ImprovedTIP5P/2018}), for which the mean solvation forces are calculated by keeping the solute position constrained and averaging over a multitude of solvent configurations. 
Note that this procedure to generate the solvation forces differs from the one proposed by Chen \textit{et al.} \cite{Chen2021MachineDynamics} as here the averaging is performed over multiple solvent configurations.
The resulting mean solvation forces are used in a next step to train the GNN.

To assess the performance of our approach, test systems were chosen that are (a) interesting for the application of implicit solvents, (b) challenging for current implicit solvents, (c) have fast kinetics, and (d) can be analysed directly without dimensionality reduction or Markov state modelling. The aforementioned criteria are met by a class of small peptides, which are able to form a salt bridge. The importance of salt bridges for the stability of proteins \cite{Meuzelaar2016InfluenceKinetics} makes them an interesting target for implicit solvent models, while previous studies by Nguyen \textit{et al.} \cite{Nguyen2013ImprovedSimulations} have shown that GB-based implicit solvents could not accurately describe them. Although the parameters of the GB-Neck2 model \cite{Nguyen2013ImprovedSimulations} have been manually adjusted such that the height of the energy barrier matched a TIP3P solvent \cite{Jorgensen1983ComparisonWater} simulation, this implicit solvent failed to reproduce other key characteristics of the system such as the position of the free-energy minimum of the salt bridge or effects attributed to a single explicit water molecule between the opened salt bridge. 

In this study, we study similar small peptides featuring a salt bridge between lysine (K, LYS) and glutamic acid (E, GLU) connected via two alanine (A, ALA) and one variable residue forming the peptides KAXAE, with X being valine (V, VAL), leucine (L, LEU), isoleucine (I, ILE), phenylalanine (F, PHE), serine (S, SER), threonine (T, THR), tyrosine (Y, TYR), and proline (P, PRO). In addition, we also test the approach on the same peptide KAAE as in Ref.~\cite{Nguyen2013ImprovedSimulations}, together with the longer variants KAAAE and KAAAAE.
The performance is compared to the state-of-the-art implicit solvent GB-Neck2 as well as explicit-solvent simulations with TIP3P and TIP5P. To explore the generalizability and transferability characteristics of the GNN, the model is challenged to simulate peptides outside of the training set. 

\section{Methods}

\subsection{Molecular Dynamics Simulations}
Starting coordinates and topologies were generated using the AmberTools21 \cite{Case2021AMBER2021} software package. The amino acids and capping groups were parametrised using the AMBER force field ff99SB-ILDN \cite{Lindorff-Larsen2010ImprovedField}. All simulations were performed using OpenMM (version 7.7.0) \cite{Eastman2017OpenMMDynamics}. For the explicit-solvent simulations, the peptides were solvated in a box of water (TIP3P or TIP5P) with a padding of \SI{1}{nm}. For all systems, an energy minimisation using the L-BFGS algorithm \cite{Liu1989OnOptimization} was performed with a tolerance of \SI{10}{\kilo \joule \per \mol \per\nano\meter}. All bonds involving hydrogens were constrained using the SETTLE \cite{Miyamoto1992Settle:Models} and CCMA  \cite{Eastman2010ConstantSimulations} algorithms for water and peptide bonds, respectively. For all simulations, Langevin dynamics were used with the LFMiddle discretization scheme \cite{Zhang2019UnifiedDynamics}. For the explicit-solvent simulations, a cutoff of \SI{1}{nm} for the long-range electrostatic interactions was used together with the particle mesh Ewald (PME) correction  \cite{Darden1993ParticleSystems}. The simulation temperature set to 300~K and a time step of \SI{2}{\femto\s} was applied. The simulations with explicit solvent or GB-Neck2 were carried out for \SI{1000}{ns}. Simulations with the GNN to describe the solvent were performed using the OpenMM-Torch package (version 0.6, \href{https://github.com/openmm/openmm-torch}{url: https://github.com/openmm/openmm-torch}) by introducing the GNN as an additional force to the vacuum force field. These simulations were carried out for \SI{30}{ns}. Note that explicit-solvent simulations with TIP3P were only performed for peptides KAPAE, KASAE, KATAE, and KAYAE.

\subsection{Generation of the Training Set}
From the explicit-solvent simulations with TIP5P, a conformer was extracted every \SI{200}{ps} for the peptides with X being apolar (i.e., KAVAE, KALAE, KAIAE, KAFAE, and KAPAE) and every \SI{100}{ps} for the peptides with X being polar (i.e., KASAE, KATAE, and KAYAE). To calculate the mean solvation forces for each conformer, the solute atoms were positionally constrained and an explicit-solvent simulation was performed for \SI{200}{ps}. The solvent-solute forces were evaluated every \SI{200}{fs} by calculating the forces for each solute atom within the explicit simulation and subtracting the forces of the same solute atom in vacuum.

\subsection{Graph Neural Networks}
Two GNN architectures were explored, sharing a three-pass neural network as core architecture. Both GNNs were applied in the subsequent simulations via a $\Delta$-learning scheme \cite{Ramakrishnan2015BigApproach}. In the first case (abbreviated as GNN+), the energies of the base model (i.e., GB-Neck2) and the GNN were summed up following a traditional $\Delta$-learning approach (the forces were then obtained via standard derivation). In the second case (abbreviated as GNN*), the functional form of the base model (GB-Neck2) was adjusted such that the Born radii are scaled by the GNN within a predefined range according to Eq.~\ref{eq:bornscaling},
\begin{equation}
    R'_i = R_i * (p + S(\phi(R,q,r_{a},r,R_{\text{cutoff}})) \cdot (1-p) \cdot 2) 
\label{eq:bornscaling}
\end{equation}
where $R_i$ is the Born radius calculated based on the Neck integral, $p$ the scaling parameter to adjust the strength of the scaling of the Born radii (i.e., 1 = no scaling applied; 0 = maximum scaling applied), $S$ the sigmoid function, and $\phi$ the function approximated by the GNN based on the Born radii $R$, the charges $q$, the atomic radii $r_a$ of all atoms, the distances $r$ between all atoms, and a cutoff radius $R_{\text{cutoff}}$. 

\begin{figure}[H]
\centering
    \includegraphics[width=\columnwidth]{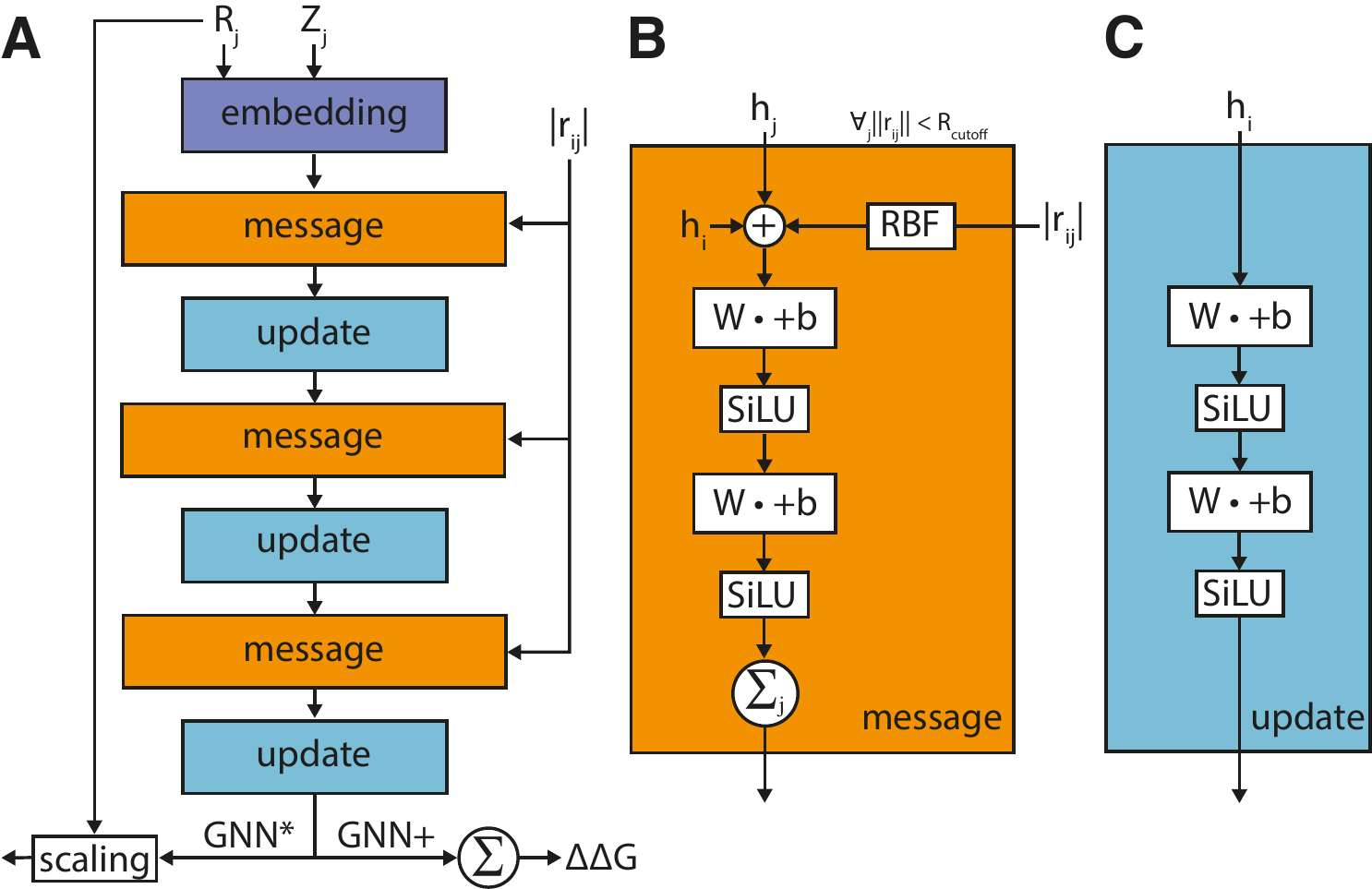}
    \caption{Schematic representation of the GNN architectures. Node-wise computations are shown in blue, while message operations are shown in orange. (\textbf{A}): Schematic representation of the computation through the GNN.
    (\textbf{B}): Message computations start by concatenating the atom features of the sending and receiving node with the distance encoded by a Bessel function (RBF), followed by a two-layer MLP with SiLU activation functions. (\textbf{C}): Node-wise computations of aggregation by summation followed by a two-layer MLP with SiLu activation functions.}\label{fig:GNNarchitecture}
\end{figure}

The GNN* and GNN+ networks share key architectural elements as both employ three passes through interaction networks that only differ in the shape of the input and output, followed by SiLU activations \cite{Elfwing2017Sigmoid-WeightedLearning}. 
One interaction network pass is characterised by a concatenation of the node features of the sending and receiving node together with the distance encoded by a radial Bessel \cite{Gasteiger2022DirectionalCoordinates} function of length 20, followed by a multi-layer perceptrons (MLP) with two layers and SiLU activation functions. As an aggregation method, the sum of all messages was taken and the node-wise computation employed again a two-layer MLP with SiLU activation functions. All hidden layers have a size of 128. As an embedding for the atoms, the GNN+ model used the partial charges and atomic radii from the force field and GBNeck2, respectively, while the GNN* incorporates additionally the calculated Born radius from the Neck integral. A schematic representation for the GNN architectures is shown in Figure~\ref{fig:GNNarchitecture}. 

The GNNs were trained on randomly selected 80~\% of the conformations, using the Adam optimiser \cite{Kingma2017Adam:Optimization} for 100 epochs with a batch size of 32 and an exponentially decaying learning rate starting at 0.001 and decaying by two orders of magnitude to 0.00001. The mean squared error (MSE) was chosen as the loss function and the samples randomly shuffled after each epoch.

\subsection{Training-Test Splits}

To assess the generalisability and transferability of the GNN, we composed different training and test splits (Table \ref{tab:TraingTestSet}). Splits 1-4 include all KAXAE peptides for training but one. In split 5, all KAXAE peptides with X = polar are used for training, while the ones with X = apolar are simulated prospectively. Split 6 is the other way around (i.e., training on X = apolar, testing on X = polar). Finally, in split 7, all peptides with five residues (except X = A) were used for training, while prospective simulations were performed for KAAE, KAAAE, and KAAAAE.

\begin{table}[H]
\caption{Training and test splits of peptides KAXAE with X being V, L, I, F, P, S, T, Y, '', A, AA (note that KAXAE with X = '' corresponds to KAAE). Peptides used for training are marked with \fcolorbox{blue!25}{blue!25}{\textcolor{blue!25}{T}}, while peptides simulated prospectively (testing) are marked with \fcolorbox{red!25}{red!25}{\textcolor{red!25}{S}}.} \label{tab:TraingTestSet}
\centering
\begin{tabular}{l | p{0.6cm} p{0.6cm} p{0.6cm} p{0.6cm} p{0.6cm} p{0.6cm} p{0.6cm} p{0.6cm} | p{0.6cm} p{0.6cm} p{0.6cm}}
\hline
Split & \hfil V &\hfil L &\hfil I &\hfil F &\hfil P &\hfil S &\hfil T &\hfil Y &\hfil '' &\hfil A &\hfil AA \\ \hline \hline
1 & \cellcolor{blue!25} & \cellcolor{blue!25} & \cellcolor{blue!25} & \cellcolor{blue!25} & \cellcolor{blue!25} & \cellcolor{red!25} & \cellcolor{blue!25} & \cellcolor{blue!25} &  &  & \\ \hline
2 & \cellcolor{blue!25} & \cellcolor{blue!25} & \cellcolor{blue!25} & \cellcolor{blue!25} & \cellcolor{blue!25} & \cellcolor{blue!25} & \cellcolor{red!25} & \cellcolor{blue!25} &  &  & \\ \hline
3 & \cellcolor{blue!25} & \cellcolor{blue!25} & \cellcolor{blue!25} & \cellcolor{blue!25} & \cellcolor{blue!25} & \cellcolor{blue!25} & \cellcolor{blue!25} & \cellcolor{red!25} &  &  & \\ \hline
4 & \cellcolor{blue!25} & \cellcolor{blue!25} & \cellcolor{blue!25} & \cellcolor{blue!25} & \cellcolor{red!25} & \cellcolor{blue!25} & \cellcolor{blue!25} & \cellcolor{blue!25} &  &  & \\ \hline
5 & \cellcolor{red!25} & \cellcolor{red!25} & \cellcolor{red!25} & \cellcolor{red!25} & \cellcolor{red!25} & \cellcolor{blue!25} & \cellcolor{blue!25} & \cellcolor{blue!25} &  &  & \\ \hline
6 & \cellcolor{blue!25} & \cellcolor{blue!25} & \cellcolor{blue!25} & \cellcolor{blue!25} & \cellcolor{blue!25} & \cellcolor{red!25} & \cellcolor{red!25} & \cellcolor{red!25} &  &  & \\ \hline
7 & \cellcolor{blue!25} & \cellcolor{blue!25} & \cellcolor{blue!25} & \cellcolor{blue!25} & \cellcolor{blue!25} & \cellcolor{blue!25} & \cellcolor{blue!25} & \cellcolor{blue!25} & \cellcolor{red!25} & \cellcolor{red!25} & \cellcolor{red!25} \\
\hline
\end{tabular}
\end{table}

\subsection{Data Analysis}
\begin{figure}[H]
\centering
    \includegraphics[width=0.85\columnwidth]{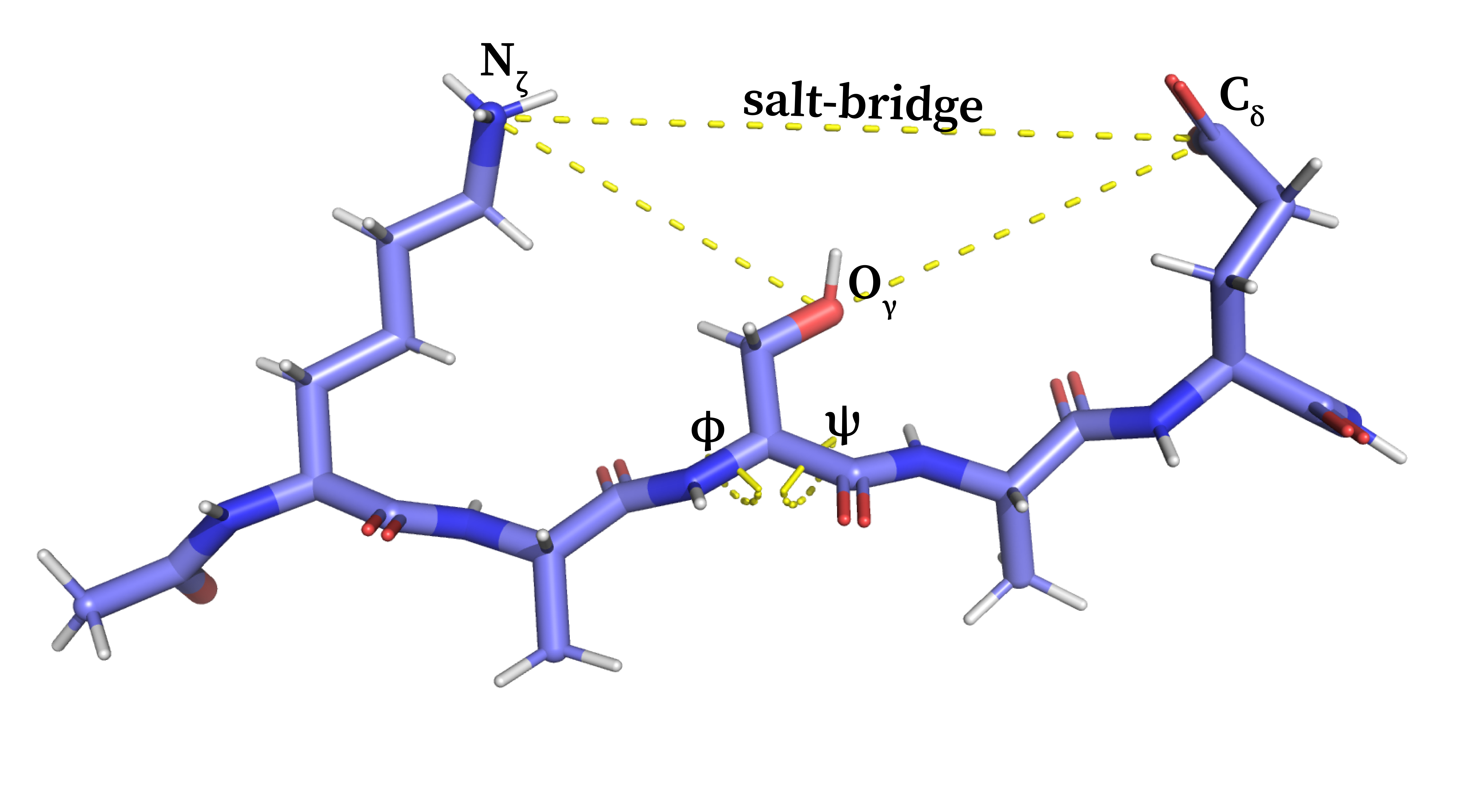}
    \caption{Structural characteristics of KASAE as example (light blue) with highlighted salt bridge, the distance SER $O_\gamma$ -- \nzf and SER $O_\gamma$ -- \cdf, and backbone torsional angles $\phi$ and $\psi$ of the central residue.}
    \label{fig:exampledistances}
\end{figure}

We used MDTraj (version 1.9.7) \cite{McGibbon2015MDTraj:Trajectories} for the analysis of the trajectories. To estimate the statistical uncertainty of the different approaches, the explicit-solvent simulations were divided into five \SI{200}{ns} blocks and analysed separately. In addition, the training and simulation of the GNN was repeated three times with different random seeds. All free-energy profiles were calculated with a Jakobian correction factor of $4 \pi r^2$ \cite{Herschbach1959MolecularProperties}.
A key feature that the ML-based implicit solvent should reproduce is the correct representation of the salt bridge between \nzf and \cdf. We have identified three main characteristics concerning the salt bridge for all test systems (Figure~\ref{fig:exampledistances}): (i) a double well in the free-energy minimum at \SI{0.33}{\nm} and \SI{0.37}{\nm} with different weights, corresponding to two different closed salt-bridge geometries, (ii) the height of the energy barrier the opening of the salt bridge at \SI{0.43}{\nm}, and (iii) a dip in the free-energy profile at \SI{0.55}{nm}, corresponding to conformations with one water molecule between \nzf and \cdf in a hydrogen-bond network. In addition to the salt bridge, the backbone dihedral angles $\phi$ and $\psi$ of the central amino acid are monitored. For polar central residues, the distance between the oxygen of the hydroxy group of the polar side chains and the \nzf and \cdf (Figure~\ref{fig:exampledistances}).

\section{Results and Discussion}

\subsection{Comparison of GNN Architectures}
\begin{figure}[H]
\centering
    \includegraphics[width=0.5\columnwidth]{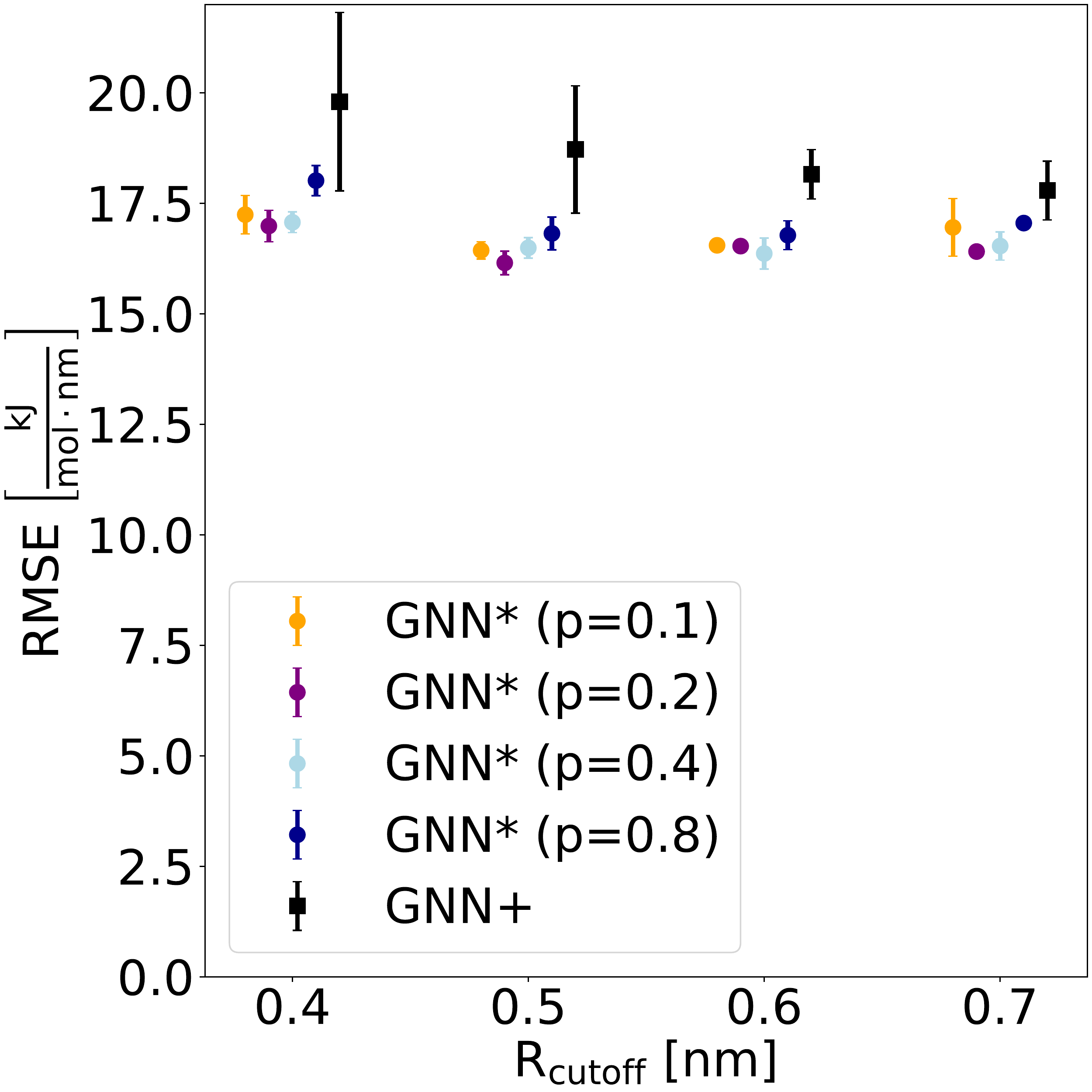}
    \caption{Comparison of the root-mean-square error (RMSE) of the forces predicted by the GNN* and GNN+ models with $\Delta$-learning for the external test set. The GNN* models with scaling parameter $p$ of 0.1, 0.2, 0.4, or 0.8 are shown as colored dots (orange, purple, light blue, and navy, respectively), while the GNN+ is shown as black squares. Statistical uncertainty is denoted by error bars.}
    \label{fig:benchmark}
\end{figure}

We explored two GNN architectures with $\Delta$-learning schemes for the ML-based implicit solvent: GNN+ and GNN*. To compare the architectures and evaluate their hyperparameters, we calculated the solvation forces of simulation snapshots in a retrospective manner. 
We considered three peptides (KAVAE, KALAE, and KASAE) for training (using a random 20 \% subset as validation set during the GNN training process), and two peptides KATAE and KAFAE as external test set. The peptides KAIAE, KAPAE and KAYAE were not included in this benchmark in order to not bias the choice of model architecture for the subsequent simulation studies. The investigated hyperparameters were the cutoff radius $R_{\text{cutoff}}$ for which the fully connected graph is constructed, as well as the scaling parameter $p$ of the GNN*, which regulates how much the Born radii of the GB-Neck2 model are allowed to change. The results of the benchmarking are summarised in Figure~\ref{fig:benchmark}. 

Interestingly, the GNN* model was found to perform significantly better than the GNN+ model in predicting the forces of the external test set over the entire range of tested cutoff radii, reaching an RMSE of \SI{16.2 \pm 0.3}{\kilo\J\per\mol\per\nm}. In addition, smaller deviations between the different random seeds were observed with the GNN* (indicated by lower standard deviations), which is a desirable feature. The effect of the scaling parameter on the GNN* models is more subtle. Values of 0.1, 0.2, and 0.4 gave essentially the same results, while the error increases slightly for a scaling parameter of 0.8. The impact of the cutoff radius $R_{\text{cutoff}}$ is also small, although $R_{\text{cutoff}}$~=~\SI{0.4}{\nano\meter} is likely too short. As longer radii (i.e. \SI{0.7}{\nano\meter}) did not improve the performance significantly but increase the computational cost, we decided to focus in the following on the GNN* architecture with cutoff radii of \SI{0.5}{\nano\meter} and \SI{0.6}{\nano\meter} together with a scaling parameter of 0.1 and 0.2.

\subsection{Prospective Molecular Dynamics Simulations}
To investigate the ability of the ML-based implicit solvent to simulate novel peptides, we composed different training and test splits (Table~\ref{tab:TraingTestSet}). First, we assessed the simulation performance of the GNN* models with with radii = \SI{0.5}{\nano\meter} or \SI{0.6}{\nano\meter} and scaling parameter = 0.1 or 0.2 on the training/test splits 1, 2, and 3 (i.e., training on all peptides except KASAE, KATAE, or KAYAE, respectively, and prospective simulation of the leftout peptide). The GNN* model with a radius of \SI{0.6}{nm} and a scaling parameter of 0.1 yielded the most stable simulation results as indicated by the smallest deviations between the different random seeds for the closed salt-bridge conformations (see Figures S1-S4 in the Supporting Information). 
The observation that models with similar performance on a retrospective test set show more different behaviour in prospective simulations is in line with findings by Fu \textit{et al.} \cite{Fu2022ForcesSimulations} and Stocker \textit{et al.} \cite{Stocker2022HowSimulations} Based on these results, the following analyses were performed using only the GNN* with a radius of \SI{0.6}{nm} and a scaling parameter of 0.1.

\subsubsection{KASAE, KATAE, or KAYAE as Test Peptide}
\begin{figure}[H]
    \centering
    \includegraphics[width=\columnwidth]{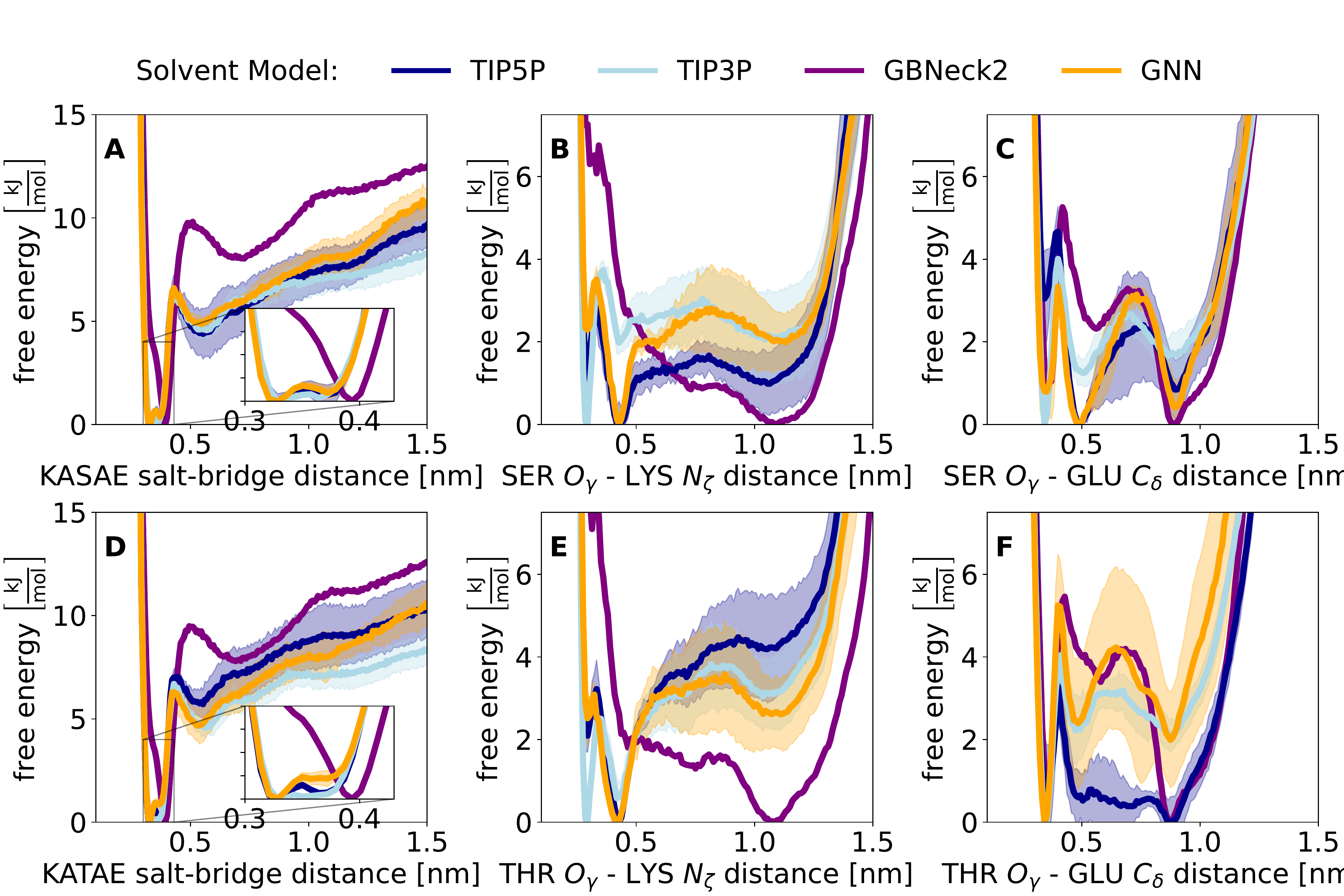}
    \caption{Comparison of the GNN* implicit solvent model (orange) with explicit TIP5P (navy blue) and TIP3P (light blue) as well as the GB-Neck2 implicit solvent (purple). Results for KASAE (split 1) are shown in the top row, results for KATAE (split 2) in the bottom row. (\textbf{A}, \textbf{D}): Free-energy profile of the salt bridge. (\textbf{B}, \textbf{E}): Distance $O_\gamma$ -- LYS $N_\zeta$. (\textbf{C}, \textbf{F}): Distance $O_\gamma$ -- GLU $C_\delta$. The shaded area indicates the statistical uncertainty of the corresponding solvent model (not shown for GB-Neck2 for clarity).}
    \label{fig:SERTHRComp}
\end{figure}

The training/test splits 1, 2, and 3 (Table~\ref{tab:TraingTestSet}) are particularly interesting as they challenge the model the most. Among the three, the simulation of KASAE (split 1) and KATAE (split 2) are expected to be easier for the model as there is a similar residue in the training set (THR versus SER). Free-energy profiles of the salt bridge, the $O_\gamma$ -- \nzf distance, and $O_\gamma$ -- \cdf distance are shown in Figure \ref{fig:SERTHRComp}. 

\begin{figure}[H]
\centering
    \includegraphics[width=0.4\textwidth]{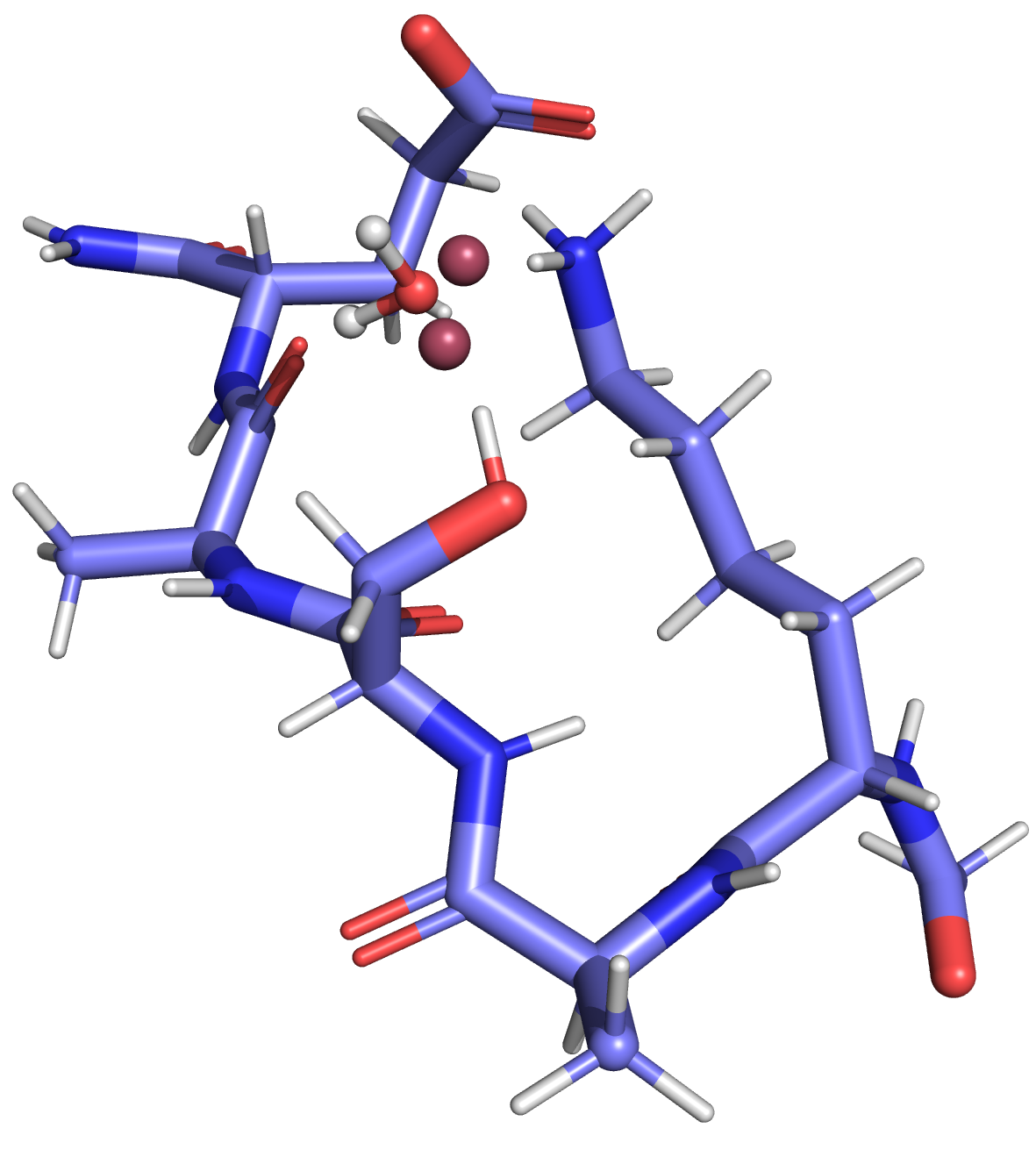}
    \caption{Example conformation of KASAE (light blue) with a SER \og -- \nzf distance of \SI{0.43}{\nm}. The TIP5P water molecule interacting with the SER \og, the \nzf, and a carbonyl of the backbone is shown with its off-site charges displayed in dark red.}
    \label{fig:SERTIP5P}
\end{figure}

The GNN* implicit solvent was able to correctly reproduce all desired properties of the salt bridge in TIP5P explicit water, featuring the correct double well in the free-energy profile at short distances, the correct energy barrier of opening for the salt bridge, and a local minima at \SI{0.55}{nm} (Figure \ref{fig:SERTHRComp}A,D). The GB-Neck2 model, on the other hand, failed to describe any of these features. Note that it has been shown by Nguyen \textit{et al.} \cite{Nguyen2013ImprovedSimulations} that the GB-Neck2 model can be tweaked to reproduce smaller barrier heights for salt-bridge opening, but the other features have not been reported for that model. 
Larger deviations can be observed for the SER/THR \og \nzf distance (Figure \ref{fig:SERTHRComp}B,E). Again, the GB-Neck2 model is not able to capture the key characteristics of the TIP5P free-energy surface, while the GNN* reproduces the minima correctly and shows also good agreement with the local minima of the direct hydrogen bond between \og and \nz at \SI{0.29}{\nm}. Interestingly, the TIP3P model shows quite different characteristics compared to TIP5P. The minima of the direct hydrogen bond is lower than the second minima at \SI{0.43}{\nm}. The latter minima contains conformations where an explicit water molecule forms a hydrogen bond network between the salt-bridge partners. Analysing similar conformations in the TIP5P solvent simulation, we could identify examples that benefit from the specific geometry of the TIP5P water representation (Figure~\ref{fig:SERTIP5P}). We hypothesise that the 'triangle' between the SER \og, the \nzf, and a carbonyl of the peptide backbone allows for a trident 'binding' site for the explicit TIP5P water molecules, thus stabilising these conformations, which is not possible with the simplified TIP3P model.

\begin{figure}[htb!]
    \centering
    \includegraphics[width=0.6\columnwidth]{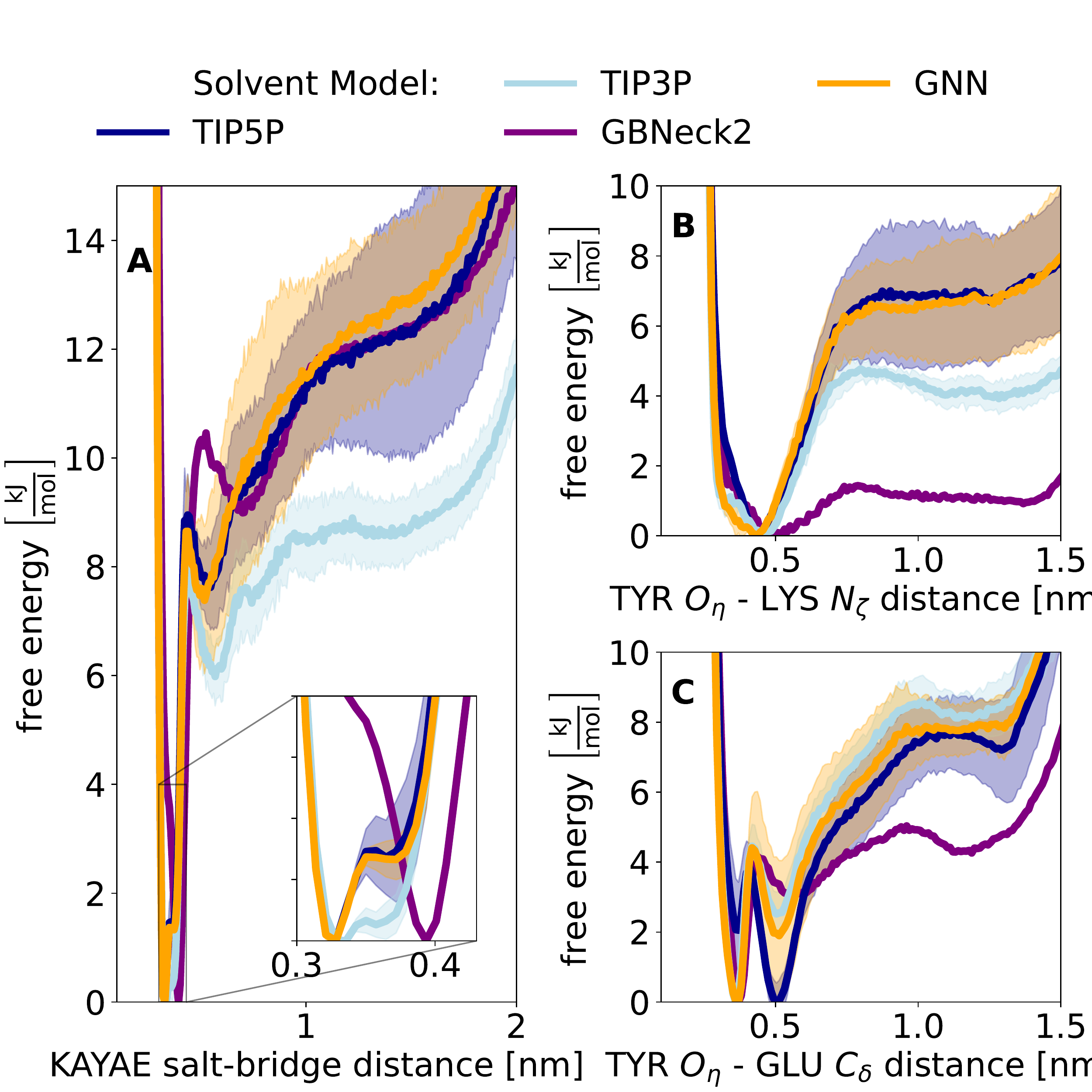}
    \caption{Comparison of the GNN* implicit solvent model (orange) with explicit TIP5P (navy blue) and TIP3P (light blue) as well as the GB-Neck2 implicit solvent (purple). Results for KAYAE (split 3) are shown. (\textbf{A}): Free-energy profile of the salt bridge. (\textbf{B}): Distance TYR $O_\eta$ -- LYS $N_\zeta$. (\textbf{C}): Distance TYR $O_\eta$ -- GLU $C_\delta$. The shaded area indicates the statistical uncertainty of the corresponding solvent model (not shown for GB-Neck2 for clarity).}
    \label{fig:TYRComp}
\end{figure}

A more challenging case is KAYAE (split 3) as none of amino acids in the training set is very similar to TYR. Thus, the model needs to learn about the solvation of the TYR side chain from  different amino acids, i.e., the learning task becomes to generalise from PHE and SER/THR to a combination of the two. If the model achieves a good accuracy for split 3, it demonstrates its transferability to peptides with increasing structural differences to the training set. The free-energy profile of the KAYAE salt bridge, and the distances TYR $O_\eta$ -- \nzf and the TYR $O_\eta$ -- \cdf are shown in Figure~\ref{fig:TYRComp}. 
Again, the GNN* model reproduces the salt-bridge free-energy profile of TIP5P very well (Figure~\ref{fig:TYRComp}A). Interestingly, the TIP3P solvent shows in this case a different behaviour than TIP5P at short distances. The double well of the salt bridge is significantly different and the barrier height (at \SI{0.43}{\nm}) is lower for TIP3P. For the distance TYR $O_\eta$ -- \nzf, the TIP5P and GNN* show the same behaviour, while the energy barrier with TIP3P is of approximately \SI{3}{\kilo\J\per\mol} lower (Figure~\ref{fig:TYRComp}B). For the distance TYR $O_\eta$ -- \cdf, the minimum with TIP3P and GNN* is the direct H-bond distance at \SI{0.37}{\nm}, while the minimum with TIP5P is at a distance of \SI{0.51}{nm}, where one explicit water interacts in a H-bond network between the salt-bridge partners (Figure~\ref{fig:TYRComp}C).

\begin{figure}[H]
    \centering
    \includegraphics[width=0.75\textwidth]{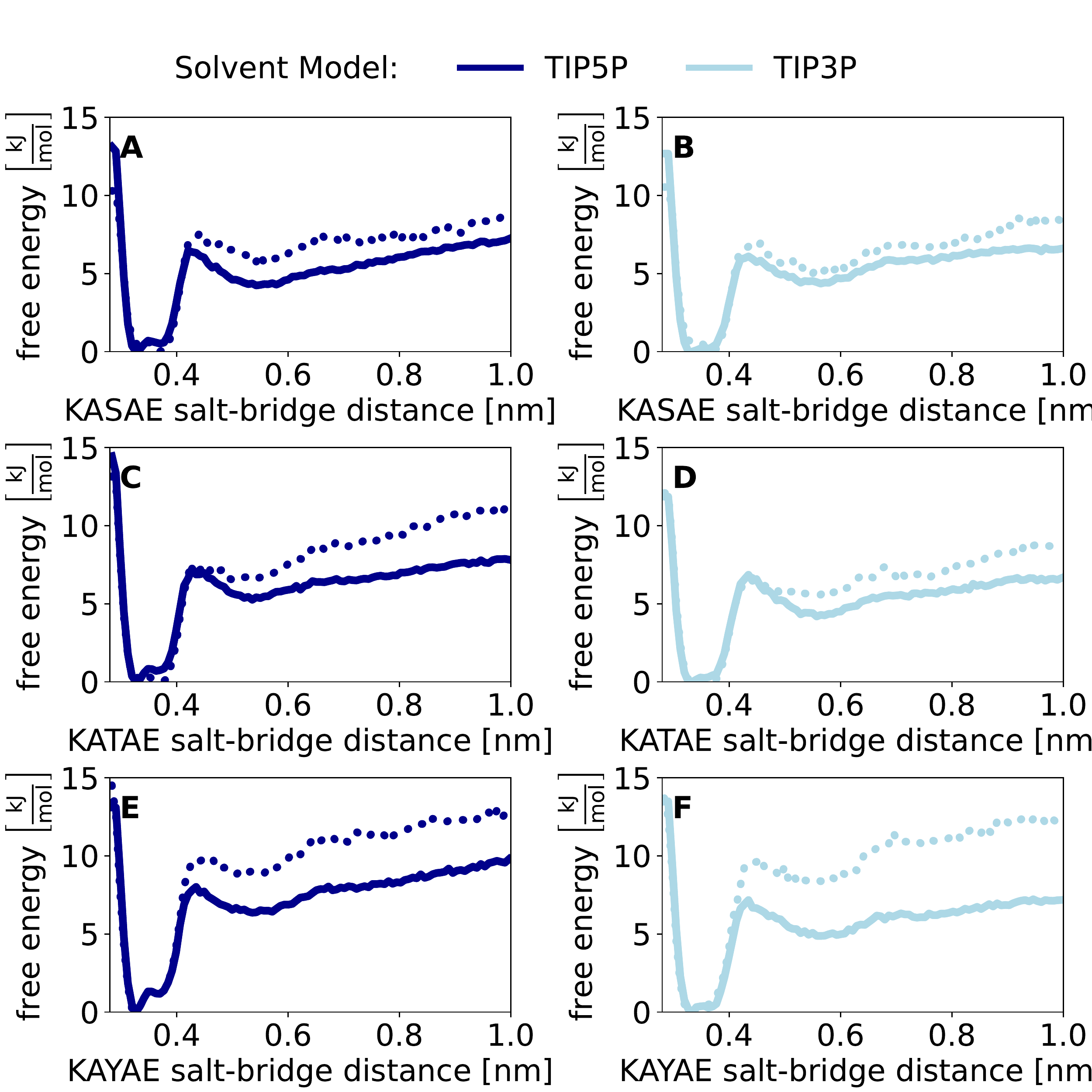}
    \caption{Free-energy profile of the salt-bridge distance for KASAE (top), KATAE (middle), and KAYAE (bottom) in the explicit-solvent simulations with TIP5P (left (\textbf{A}, \textbf{C}, \textbf{E}) navy blue) and TIP3P (right (\textbf{B}, \textbf{D}, \textbf{F}) light blue). The dotted line corresponds to conformations with the distance between $O$ of the hydroxy group and LYS $N_\zeta$ $<$~\SI{0.60}{nm} and the solid line to conformations with the distance between $O$ of the hydroxy group and LYS $N_\zeta$ $>$~\SI{0.60}{nm}.}
    \label{fig:bardiffTIP5PTIP3Pcomp}
\end{figure}

While the free-energy profiles of the salt bridge were similar for TIP5P and TIP3P for KASAE and KATAE, they differed for KAYAE. Therefore, we investigated this observation further. The difference in barrier height for opening the salt bridge (Figure~\ref{fig:TYRComp}A) could be due to different preferences for the distance TYR $O_\eta$ -- \nzf. In Figure~\ref{fig:bardiffTIP5PTIP3Pcomp}, the free-energy profile of the salt bridge for KASAE, KATAE, and KAYAE was calculated once for conformations with the distance between $O$ of the hydroxy group and \nzf $<$~\SI{0.6}{nm} and once for those with this distance $>$~\SI{0.6}{nm}. Intriguingly, the energy barrier to open the salt bridge of KAYAE is higher when TYR $O_\eta$ interacts with the \nzf (e.g., either via a direct hydrogen bond or mediated by a water molecule) in both TIP5P and TIP3P water. For KASAE and KATAE, this is not the case. From these findings, it follows that because TIP5P and GNN* favour conformations of KAYAE with a distance TYR $O_\eta$ -- \nzf $<$~\SI{0.6}{\nano\m} more than TIP3P (Figure~\ref{fig:TYRComp}B), the energy barrier of the salt bridge is higher than in TIP3P.

\subsubsection{Proline as Special Case: KAPAE}
\begin{figure}[H]
\centering
\includegraphics[width=0.6\textwidth]{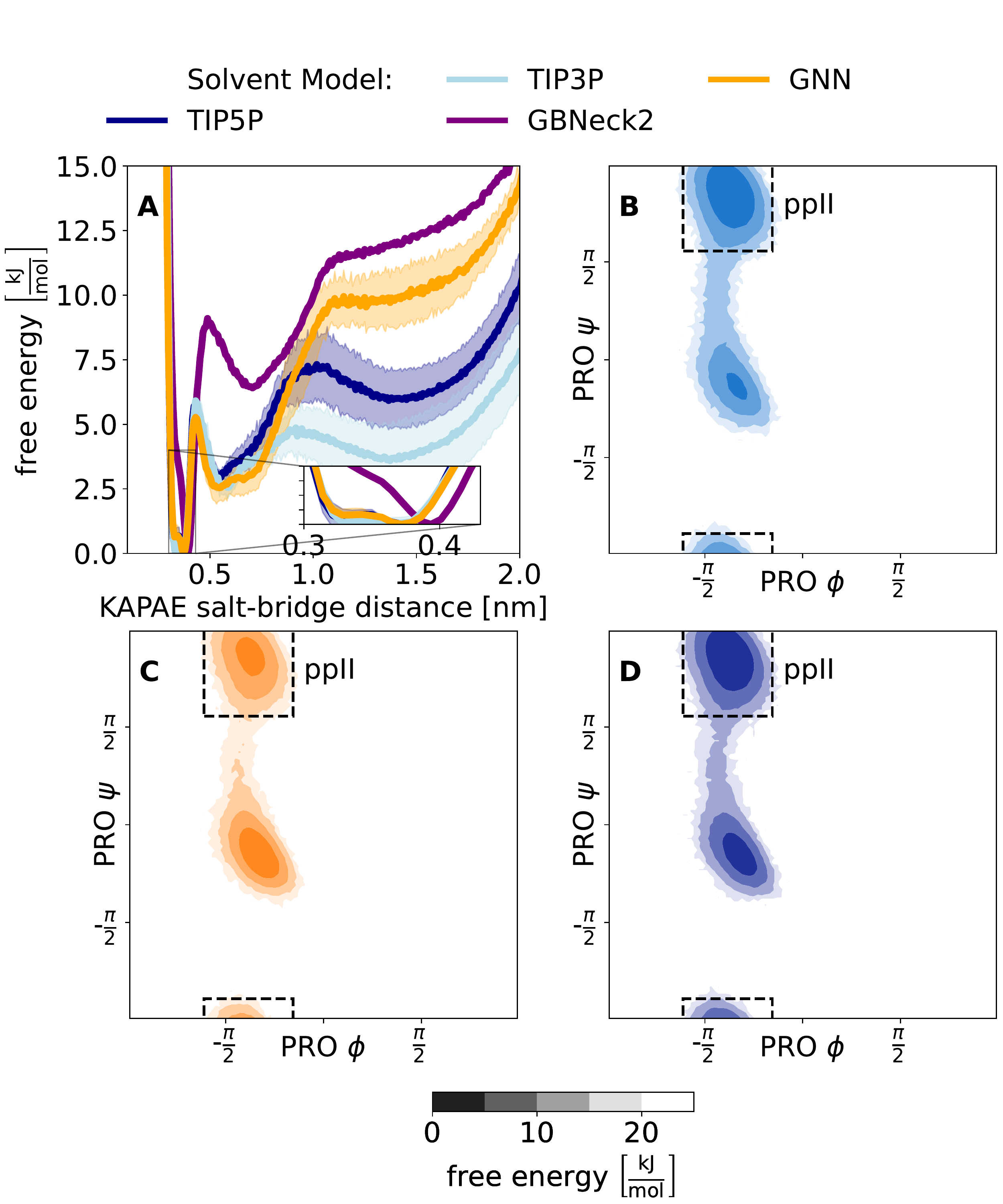}
\caption{Comparison of the GNN* implicit solvent model (orange) with explicit TIP5P (navy blue) and TIP3P (light blue) as well as the GB-Neck2 implicit solvent (purple). Results for KAPAE (split 4) are shown. (\textbf{A}): Free-energy profile of the salt bridge. The shaded area indicates the statistical uncertainty of the corresponding solvent model (not shown for the GB-Neck2 implicit solvent for clarity). (\textbf{B}): Ramachandran plot of PRO with GB-Neck2. (\textbf{C}): Ramachandran plot of PRO with GNN*. (\textbf{D}): Ramachandran plot of PRO with TIP5P.  The polyproline II state is highlighted in the Ramachandran plots by a dashed black line.}\label{sfig:PROcomp}
\end{figure}

An interesting case is PRO as central residue (split 4 in Table \ref{tab:TraingTestSet}). Proline has a different Ramachandran plot than the other amino acids and is therefore possibly challenging to describe correctly by the GNN* approach. As can be seen in Figure~\ref{sfig:PROcomp}A, the free-energy profile of the salt bridge is reproduced well, although deviations from TIP5P occur at long distances. Interestingly, at these long distances also TIP3P and TIP5P disagree with each other. While TIP3P yields lower free energies for the open state compared to TIP5P, the GNN* predicts even higher free energies. As PRO is in the middle of the peptide, the sampling of its backbone dihedral angles will influence the long salt-bridge distances. The Ramachandran plots for PRO are shown in Figure~\ref{sfig:PROcomp}B-D for TIP3P, GNN*, and TIP5P. The main difference is in the population of the polyproline II state, which is overestimated with TIP3P and underestimated with GNN* compared to TIP5P (see also Figure~S5 in the Supporting Information). It is therefore important to include PRO in the training set to ensure proper sampling of its backbone conformations.

\subsubsection{Apolar Versus Polar Residues}
\begin{figure}[H]
    \centering
    \includegraphics[width=0.6\textwidth]{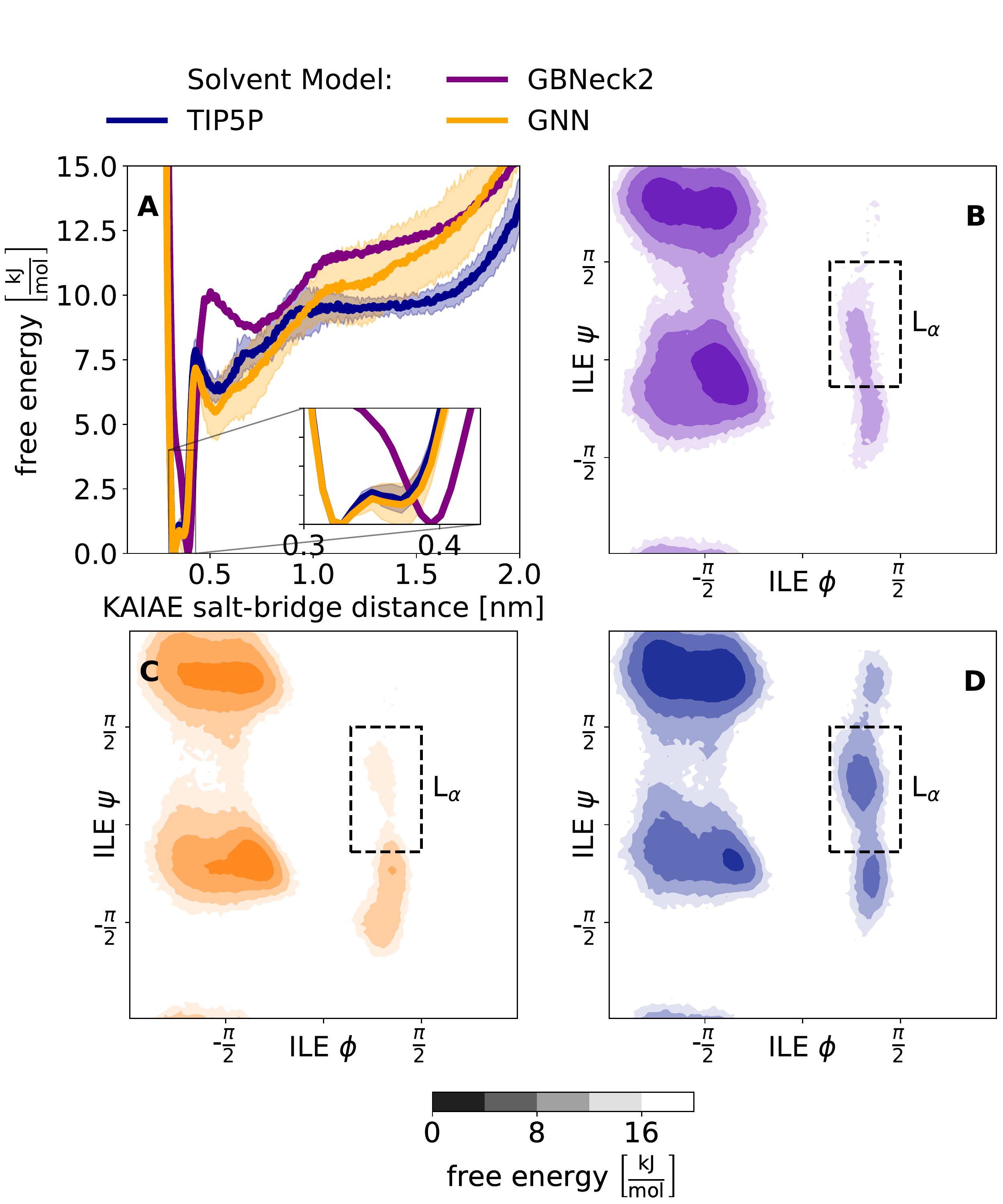}
    \caption{Comparison of the GNN* implicit solvent model (orange) with explicit TIP5P (navy blue) and TIP3P (light blue) as well as the GB-Neck2 implicit solvent (purple). Results for KAIAE (split 5) are shown. (\textbf{A}): Free-energy profile of the salt bridge. The shaded area indicates the statistical uncertainty of the corresponding solvent model (not shown for the GB-Neck2 implicit solvent for clarity). (\textbf{B}): Ramachandran plot of ILE with GB-Neck2. (\textbf{C}): Ramachandran plot of ILE with GNN*. (\textbf{D}): Ramachandran plot of ILE with TIP5P. The $\mathrm{L}_\alpha$ state is highlighted in the Ramachandran plots by a dashed black line.}
    \label{fig:polarnonpolar_kaiae}
\end{figure}

The results above demonstrated that the GNN* model is able of reproduce key characteristics of explicit-water simulations of peptides different from the training set. Next, we investigated how different training set compositions influenced the simulation performance of the model. We therefore created two training sets (splits 5 and 6 in Table~\ref{tab:TraingTestSet}) by dividing all peptides into two excluding subsets: (1) central residue has a polar side chain (i.e., KASAE, KATAE, and KAYAE), and (2) central residue has a apolar side chain (i.e., KAVAE, KAIAE, KALAE, KAFAE, and KAPAE). In split 5, training was carried out with the first group and prospective simulations were performed for the second group. As can be seen in Figure~\ref{fig:polarnonpolar_kaiae} for KAIAE, the generalization from the peptides with a polar central residue to those with an apolar one is good. For all tested peptides, the GNN* model is able to reproduce the free-energy profile of the salt bridge, the double well minima, the height of the energy barrier for opening, and the first dip of the reference TIP5P simulation. The corresponding results for KAVAE, KALAE, KAFAE, and KAPAE are shown in Figures~S6-S9 in the Supporting Information. 

To probe the conformational sampling of the central residue in more detail, we compared its Ramachandran plot for the different solvent descriptions. With the exception of the $\mathrm{L}_\alpha$ state, the Ramachandran plots with GNN* and TIP5P agree well. Note that the transition into the $\mathrm{L}_\alpha$ state is a rare event. In the TIP5P reference simulations of KAIAE, this state is sampled in only one of the five \SI{200}{ns} blocks (see Figures S10-S14 in the Supporting Information). 
The differences in the population of the $\mathrm{L}_\alpha$ state between GNN* and TIP5P may therefore stem from finite sampling effects.

The inverse, i.e., training on peptides with an apolar central residue and testing on peptides with a polar central residue (split 6), is more challenging. The GNN* model was still superior to the GB-Neck2 implicit solvent in reproducing key characteristics of the TIP5P reference simulations, however, deviations were observed for the interactions of the polar central residue with the salt-bridge partners (i.e., distance between $O$ of the hydroxy group and \nzf/\cdf) (see Figures S15-S17 in the Supporting Information). 
These results indicate that the extent to which the GNN* model can generalise from the training set is limited to similar functional groups. For instance, if no hydroxy group is present in the training set, the ability of the model to represent its interactions is limited. On the other hand, the TYR case demonstrates that the model is able to generalise from a hydroxy group in SER/THR to one in a different local environment. Taken together, these findings suggest that it is important for a generally applicable GNN* implicit-solvent approach to include all functional groups in the training set, but that the model does not have to have seen the complete molecule for good performance.

\subsubsection{Varying the Length of the Peptide}
\begin{figure}[H]
\centering
    \includegraphics[width=\columnwidth]{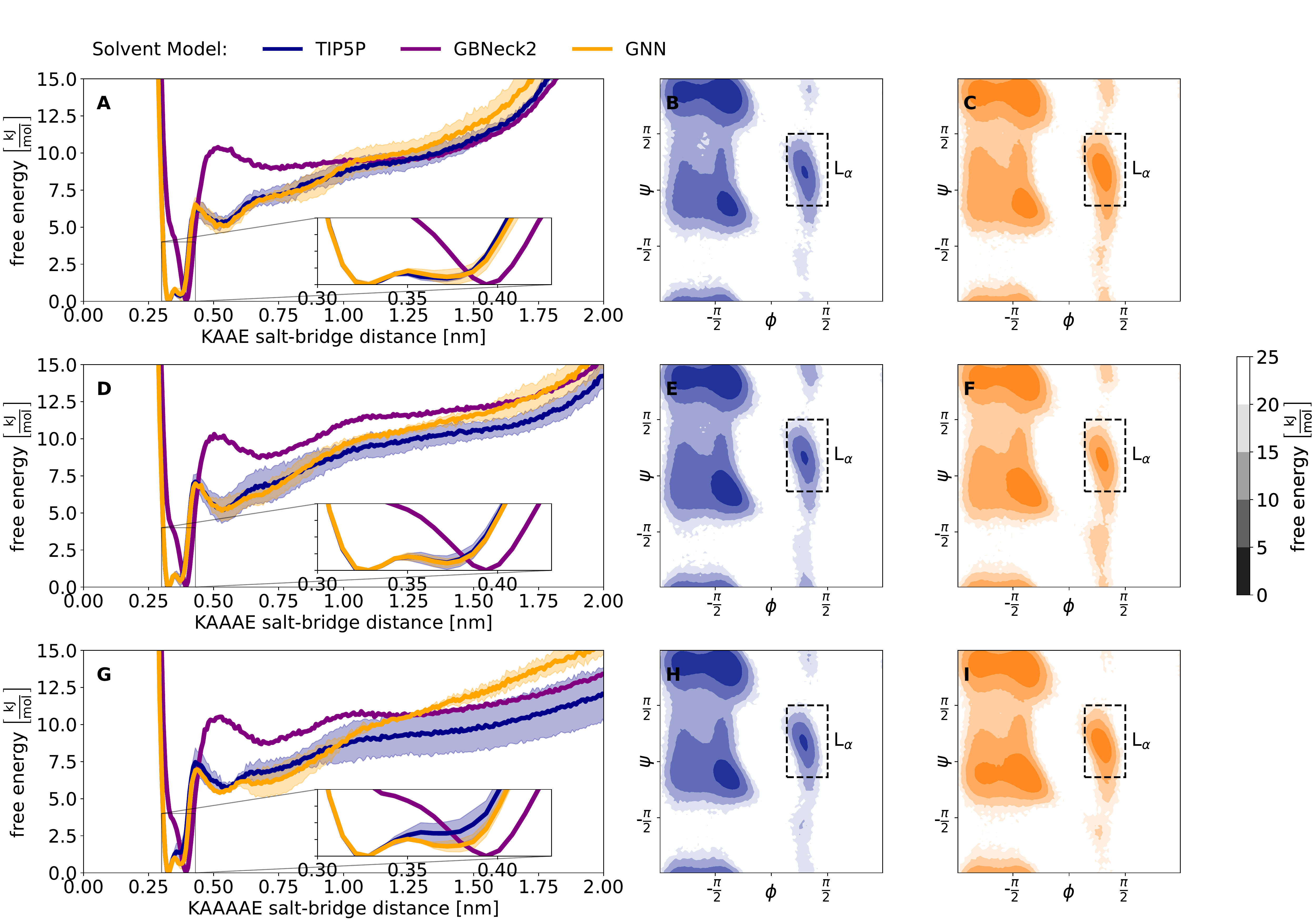}
    \caption{Comparison of the GNN* implicit solvent model (orange) with explicit TIP5P (navy blue) as well as the GB-Neck2 implicit solvent (purple). Results for KAAE (top), KAAAE (middle), and KAAAAE (bottom) are shown (split 7). (\textbf{A}, \textbf{D}, \textbf{G}): Free-energy profile of the salt bridge. The shaded area indicates the statistical uncertainty of the corresponding solvent model (not shown for the GB-Neck2 implicit solvent for clarity). (\textbf{B}, \textbf{E}, \textbf{H}): Combined Ramachandran plot of all ALA residues with TIP5P. (\textbf{C}, \textbf{F}, \textbf{I}): Combined Ramachandran plot of all ALA residues with GNN*.  The $\mathrm{L}_\alpha$ state is highlighted by a dashed black line.}\label{fig:difala_kaaaae}
\end{figure}

Finally, we investigated whether the GNN* model is able to generalise to larger or smaller peptides by removing the middle amino acid or instead inserting an extra ALA residue, i.e., peptides KAAE and KAAAAE (split 7 in Table~\ref{tab:TraingTestSet}). In addition, we included KAAAE (same size) in the test set for comparison. The resulting free-energy profiles and Ramachandran plots for KAAE, KAAAE, and KAAAAE are shown in Figure~\ref{fig:difala_kaaaae}. 
For all three peptide lengths, the GNN* is able to reproduce the free-energy profile of the salt bridge of the TIP5P reference simulation for short distances (i.e., < \SI{1}{\nano\meter}), including the double well, the height of the energy barrier, and the first dip. For the KAAE and KAAAE case also the long range behavior matches the TIP5P simulation to a high degree. Only for the KAAAAE, a deviation between GNN* and TIP5P is observed at longer distances (i.e., > \SI{1}{\nano\meter}), which could highlight a potential weak point of the GNN. While generalization to shorter peptides works well, longer peptides require either the inclusion in the training set or the introduction of a long-range correction in order to describe the elongated conformations accurately. As discussed above, differences in the population of the $\mathrm{L}_\alpha$ state may likely be finite sampling effects.

\subsubsection{Timings}
One major advantage of standard implicit solvent models is that they are much faster to compute than explicit solvent molecules. When employing GNNs for this task, the computational costs are currently still too high. 
Using a desktop PC with an Intel(R) Xeon(R) W-1270P CPU with a clock rate of 3.80GHz and a NVIDIA(R) Quadro(R) P2200 GPU, approximately \SI{46}{\nano\s\per\day} of the peptide KASAE could be obtained with our proof-of-concept implementation of the GNN implicit solvent, whereas approximately \SI{200}{\nano\s\per\day} were reached with explicit TIP5P simulations. Similar observations were made in Ref.~\cite{Doerr2021TorchMD:Simulations} for classical force-field terms. The slower speed of GNNs represents a major challenge for its application to replace explicit solvent simulations. However, this is primarily a technical issue and not a fundamental limitation. While the TIP5P explicit simulation is highly optimised, our GNN implementation is not yet. Currently, the GNN is evaluated on the GPU while the classical forces are evaluated on the CPU leading to high communication cost and low utilisation of the GPU. Recently, two approaches have been reported to increase dramatically the speed of NN potentials. The first option is the optimisation of the operations of the GNN to better suite the application of MD simulations \cite{NNPMM2022}. The second option involves batching of multiple simulations that run on one GPU in parallel \cite{Chen2021MachineDynamics}. Both approaches have been shown to bring the speed of NN potentials on par with their classical counterparts. In this work, we focused on providing a conceptual proof that developing an ML based transferable implicit solvent is possible. Improving the computational performance of the implementation is part of future work to develop a practically usable GNN implicit solvent.

\section{Conclusion}
In this work, we have developed a GNN-based implicit solvent that can be trained on a set of peptides and used to prospectively simulate different ones. The GNN* model is based on the GB-Neck2 implicit solvent with a $\Delta$-learning scheme. 
To validate our approach, we have chosen a traditionally hard problem for implicit-solvent models where the local effects of explicit water molecules play a key role: the free-energy profile of a salt bridge. Here, the salt bridge is formed by peptides with the composition KAXAE, where X can be varied. We could demonstrate that the GNN* implicit solvent was able to reproduce the key characteristics of the reference explicit-solvent simulations with TIP5P,  matching or surpassing the accuracy of explicit-solvent simulations with the simpler TIP3P water model. With different training/test splits, we assessed the ability of the GNN* model to generalise to unseen amino acids and varying peptide length. Overall, we found that the model has a high transferability as long as all functional groups are represented in the training set. For instance, if an aliphatic hydroxy group (SER or THR) is in the training set, it is sufficient for the model to correctly describe the aromatic hydroxy group of TYR. These findings are encouraging as they suggest that the training set for a globally applicable ML-based implicit solvent model may not need to be extremely large but ``only'' contain all necessary functional groups.
The results of this work present an important step towards the development of such a model, capable of replacing explicit-solvent simulations to an increasing degree.


\section*{Data and Software Availability}

The code used to generate the results of this study is available at
\url{https://github.com/rinikerlab/GNNImplicitSolvent}. Topologies, starting structures, examples for the performed analysis, and the training data are provided at the ETH Research Collection (\url{https://doi.org/10.3929/ethz-b-000599309}). Complete trajectories are available from the corresponding author upon reasonable request.

\section*{Acknowledgments}
The authors gratefully acknowledge financial support by ETH Zurich (Research Grant no. ETH-50 21-1). The authors thank Moritz Th\"urlemann for helpful discussions.

\printbibliography

\end{document}